\documentclass[preprint]{revtex4}

\begin{document}

\preprint{Preprint}

\title{An Insurance-Led Response to Climate Change}

\author{Anthony J. Webster \& Richard H. Clarke} 

\email{anthony.webster@oxfordalumni.org}

\date{\today}

\begin{abstract}
Climate change is widely expected to increase weather related damage
and  the insurance claims that result from it.
This will increase insurance premiums in a way that is independent of
a customer's contribution to the causes of climate change.
Insurance provides a financial mechanism that mitigates some of the
consequences of climate change, allowing damage from increasingly
large or frequent events to be repaired.
We observe that the insurance industry could offset any increases in 
premiums due to climate change through a levy on insurance
premiums for fossil-fuel producers for example, without needing   
government intervention or a new tax. 
We argue that an insurance-led levy or government-led carbon tax must
acknowledge a modern industry's present carbon emissions and its
fossil-fuel heritage (its ``carbon inheritance''), that is, it should 
recognise that fossil-fuel driven industrial growth has provided the
innovations and conditions needed for them to exist and develop. 
These industries have inherited the benefits of past emissions, and
we argue that they should also inherit some of the cost of those past 
emissions, 
as manifested through any statistical increase in weather-related
damage due to climate change.  
A carbon-intensity weighted tax or levy on energy production is one
mechanism that would 
recognise a modern industry's carbon inheritance, 
through increased energy costs of manufacturing and technology. 
It can be weighted to encourage the least polluting industries, and to
discourage unnecessarily carbon-intense industries. 
The cost increases would initially be small, and will require an event
attribution methodology to determine their size.  
A tax or levy can be phased in as the science of event attribution
becomes sufficiently robust for each claim type, with the latter
avoiding the need for government-led taxes, agreements, and
intervention, through a global  insurance-led response to climate
change. 
\end{abstract}

\maketitle

The insurance industry is a multi-trillion dollar business with a 2012
world premium volume of \$4 trillion US dollars \cite{PSI}, roughly
5-6\% of world GDP (that was \$71 trillion in 2012 \cite{WDI}), with
\$24 trillion of global assets under management \cite{PSI}.
For the payment of an insurance premium, consumers and businesses are
protected from the cost of rare and unpredictable events that they may
not otherwise be able to afford.
In this respect insurance provides a useful social function \cite{ISocial}.
Climate change is widely expected to increase the frequency and size
of insurance claims above that due to economic growth  
\cite{Mills,GAssStat}. 
Increased insurance premiums will be required to meet any increased
claim costs due to climate change, making insurance less affordable,
especially for the poorest and often most vulnerable people in
developing countries \cite{GAssStat,IPCC}.  
This is unfortunate, because insurance provides a re-distributive
financial protection against erratic climate events, helping to offset
some of the consequences of climate change.
The primary anthropogenic contribution to climate change is from  
cumulative CO$_2$ emissions \cite{IPCC}, but insurance costs do not
reflect a consumer's individual contribution to them. 
For example, a developing country's emissions {\it per capita} are a small
fraction of those in developed countries such as those in Europe
\cite{EPerCap}.
Arguably it would be fairer if the climate change related increases 
to insurance costs were paid for by the polluter, either directly or
indirectly through a carbon tax for example \cite{CTax0,CTax1}.
In 2015 six European Union oil and gas companies engaged with the
United Nations to develop such a tax \cite{NYTimes}. 
However, the insurance industry is in an almost unique position of being able to
reclaim these costs directly itself through increased insurance costs
on e.g. fossil-fuel producing industries.
This would require a degree of co-operation with appropriate fiscal
oversight and agreement between insurance companies to start such a
scheme, but there is a clear incentive to do so, and a strong
scientific and ethical case to support it.
The insurance industry is well organised, with international
organisations such as the Global Federation of Insurance Associations
(GFIA), the Association of British Insurers (ABI), and
The Geneva Association.
It has already demonstrated a willingness to take the lead in tackling
climate risk, with 66 CEOs of the world's leading insurance companies
in 2014 endorsing the Principles for Sustainable Insurance (``PSI''),
a set of guiding principles on the role insurance can play to tackle
climate change related risk \cite{PSI}.
At present there are few cases \cite{Otto,Pielke,Webster} where there
is a clearly demonstrable link between increased claim sizes and
anthropogenic climate change.  
Therefore there is an opportunity to gradually phase in the costs of
attribution, by allocating increased costs to climate change for each
claim type when the scientific case is clear. 
The acceptance of, and agreement about, what proportion of a claim 
might be 
attributed to climate change could involve some form of international
certification or enhanced peer review, for example through
the EUCLEIA EC project.
Alternatively the decision could be made directly by insurance
companies, in which case a degree of regulation might be required to
ensure that the best and most up-to-date scientific knowledge is used.
The agreed attribution costs could subsequently be used to  define a
carbon price. 
Event attribution presently takes much effort and intense peer-review.
The results are likely to be clearest for temperature-related risks
such as heat waves and flooding, but remain challenging for
``singularity" type events whose appearance and impact are very
difficult to predict, such as hurricanes and tornadoes for example.
But the picture is changing as event attribution science progresses,
and within ten years we expect to see widespread availability of
near real-time event attribution systems \cite{Hannant}.
Ultimately these mechanisms can transfer the climate-change induced
insurance costs to the fossil fuel industry, and subsequently to the
consumers of fossil fuels (and chemicals, plastics, etc), with higher
total contributions from those who produce the most CO$_2$.
However, to allow carbon emissions to be reduced in a humanitarian way 
in the long-term, alternatives to fossil fuels will need to be
provided.
A price disincentive alone is insufficient to prevent fossil-fuel use,
although it is argued that increased prices have successfully led to
the development of alternative energy sources in the past, by
providing an incentive for innovations and their development
\cite{Lacalle}. 
The funds raised from the proposed levy would be substantial, and 
there is the option of using them to accelerate the transformation
away from the use of fossil-fuels.  
If we chose to prioritise the outcome for both present {\sl and} future
populations, as opposed to solely the best outcome for today, then the
optimum course of action might include some diversion of funds for
mitigation or the development of low-emission alternatives.  
In the decades ahead the effects of climate change are expected to
become more pronounced, and even if emissions are reduced their
cumulative total will remain a palpable threat.
However, if fossil fuel emissions are successfully reduced then the 
scheme will demand contributions from a shrinking number of fossil
fuel producers. 
Without the primary producers to pay the levy, who could, or should,
cover the ongoing  future climate change costs? 
Presuming that we will continue to need abundant energy in the future,
and that the levy reflects energy production in addition to carbon
intensity, then revenues would be protected in a low- or post-carbon
world by an increasingly large proportion being supplied by non-fossil
fuel energy producers.
An energy tax or levy would appear to be sufficient to retain income
in the long-term, but why would it be justifiable for the resulting
price increases to be shared by all consumers? 
We propose that a levy needs to reflect not just present carbon
emissions, but also to recognise that: 

\begin{itemize}

\item[ (I) ]{No business or industry is wholly carbon neutral, and
    either indirectly or directly produces carbon emissions in their
    operation. For example, even subsistence farming today uses tools
    produced with carbon emissions. Energy use is one simple proxy for
    indirect carbon emissions.}

\item[ (II) ]{Independent of a business or industry's present energy
    use and carbon emissions (either indirect or direct), a levy
    should recognise that modern civilisations have a fossil-fuel
    heritage and a ``carbon 
    inheritance'' - that is, they are the result of over 250 years of
    fossil-fuel driven industrial growth that has provided the
    innovations and conditions needed for their existence and
    functioning - they have inherited the benefits of fossil-fuels,
    they should also inherit some of the resulting ``carbon debt''.}

\end{itemize}

Acknowledging our fossil-fuel heritage is not about allocating blame
(financial or otherwise) to historic emissions, it is about
recognising our carbon inheritance, and paying for the benefits 
that it brings.  
There are many positive outcomes associated with past emissions, 
with increased standards of living and improved health-care for
example. 
These benefits are being felt increasingly widely throughout the
developed and developing world, and by an increasingly large population. 
However it is necessary to recognise the long period of
fossil-fuel driven research and development that is embodied in every 
modern technological product    
- today's consumer benefits from past emissions.
Those emissions allowed modern products and technologies to be
innovated and developed, and this should (in principle) be reflected
in a small price increase to pay for the climate change costs that we
have also inherited. 
A financial recognition of modern products' carbon inheritance  
at the point of purchase or use, seems necessary to provide the 
long-term revenues needed for a successful response to climate change. 
The costs associated with our carbon inheritance could be determined in a 
number of ways, with various degrees of complexity.  
One approach would be to discern as fairly and accurately as possible
the likely carbon emissions leading to the long-term
development and production of a given product. 
This seems complex and intractable at present, but one could imagine
scenarios where every product is digitally identified, and all
contributions to it can be identified, traced back, and appropriately
priced. 
Another approach might be to reflect the time from the industrial
revolution's start to the beginning of a particular industry, or to
the registration of a business. 
An appealing option is to simply take energy-use as a proxy
for technology use, through a tax or insurance-levy on energy
production, and trust that it is a fair measure for most cases.  
This would go some way to reflecting our carbon inheritance and to
discourage profligate energy use. 
It is also comparatively simple, and seems to be sufficient to ensure a
long-term revenue stream to combat the effects of climate change. 
Clearly the different options outlined above require further
consideration; our
primary aim here is to highlight the ingredients that we regard as
necessary and (possibly) sufficient for a workable financial solution. 
Using fossil-fuel is like using cheap credit - it is cheap and easily
available, but has longer-term costs that will be paid sooner or
later. 
Unfortunately the slow natural removal of excess atmospheric CO$_2$ is
causing our ``carbon debt'' to increase faster than it is repaid
\cite{Clarke}.  
Therefore a disincentive for carbon emissions also seems to be
essential if we wish to minimise the rate of climate change, and the 
consequences that accrue from it. 
In the short term a tax or levy needs to be weighted so that 
it is higher for industries with the highest rate of emissions, to
encourage the use and development of lower carbon technologies and to
discourage ``embedded carbon''. 
In the longer term the costs need to smoothly transfer to low-carbon 
industries as fossil-fuel industries are replaced. 
Simple, workable schemes to achieve both of these aims are not
difficult to imagine, one example is in the Appendix.
All of these schemes avoid penalising the development of low-carbon
alternatives; only when they are the dominant producer will they
absorb a significant proportion of total climate change costs. 
These costs are likely to remain a comparatively small fraction of total
energy cost.  
Many of the arguments put forward in this paper can by extension, be adapted
to non-insurance events, that are estimated 
to contribute three times more loss \cite{Clarke}.
This would take the scheme into the realm of national governments and
their disaster relief agencies. 
Another possibility is to acknowledge the contribution to global
warming from methane emissions in the agricultural sector, again
through either a carbon tax or insurance-led levy. 
In principle we support the intentions of both suggestions, but they
add extra complexity and are a digression from the main topics that we 
wish to highlight here. 
We have also avoided a discussion of the more general impacts of
climate change on biodiversity and wildlife, and have focused 
on human interests.
Some final remarks on the rationale for, and operation of,
an insurance-led scheme. 
In principle insurance acts as an individual risk-transfer mechanism,
but as noted earlier the benefits are in practice far broader
\cite{ISocial}. 
Maintaining affordable insurance is in the interest of insurance
companies and their customers, as is a world economy that is
successfully adapting to climate change.
What we aim to highlight here are the benefits to both the insurance
industry and its customers of offsetting any  increases in claims due
to climate change, through either a carbon tax or insurance-led levy 
on the greatest contributors to climate change and on the greatest
beneficiaries from past carbon emissions. 
The implementation of an insurance-led scheme would probably require
international co-operation among insurance companies, but as discussed
earlier, the insurance industry is well organised and could arrange
this without the inter-governmental agreements needed for a globally
operating carbon tax and the effective redistribution of its proceeds.
Because insurance companies often serve 
customers from different geographic areas, an insurance-led
levy may need funds to be centrally collected and then redistributed.
In that case it could be argued: {\sl Everyone has benefited from
our carbon inheritance, albeit to different degrees, so why not simply
use peoples' consumption as a proxy for how much they have benefited?
Then collect and redistribute the income centrally as either insurance
subsidies, or for e.g. flood defences, or to help develop alternatives
to fossil fuels?} 
There are advantages to this approach \cite{Clarke}. 
For example, a simpler attribution mechanism may be sufficient to determine
the climate change cost, and some of the transaction costs associated
with insurance markets could also be avoided, albeit replaced by 
administration costs for the scheme.   
It would also avoid any conflicts of interest that may arise between
insurers who invest in fossil fuel industries for example. 
Clarke \cite{Clarke} develops the case for a carbon tax to raise
revenues that are 
administered and re-distributed by one or more of
the insurance industry, the UN, or the World Bank. 
However, this scheme is likely to require a
flow of payments from wealthier countries to those
that suffer the worst consequences of climate change. 
This is fine in principle, but will require politicians and
electorates agreement, independent of their economic and policital
circumstances.   
An insurance-led approach has the advantage that it is in the
industry's interest to successfully mitigate the increased insurance
costs due to  climate change, and any flows of capital between
companies (to offset insurance costs), can be arranged to avoid
damaging any given company's profits.  
As noted earlier, it is presently thought that about one third of all
climate change related losses are insurance related.  
We would encourage any scheme that acts to offset the consequences of
climate change in an equitable way, especially one that can continue
to operate beyond the fossil-fuel era, possibly through a mechanism
that recognises our carbon inheritance and debt as suggested here. 
To summarise, whereas a government-led carbon tax could be used to 
offset any increases in insurance premiums that are (shown to be) 
caused by climate change, the insurance industry is in a unique position of 
being able to directly offset any increases in premiums through increased 
premiums on fossil-fuel and other energy producers, for example.  
The insurance industry is sufficiently large and well organised to be
able to internationally implement either scheme.  
The advantage of an insurance-led levy is that it does not require 
negotiations with binding agreements that may be more deterimental to
some countries than others (although 
negotiations will be required within the insurance industry). 
There is a clear economic, scientific, and ethical case for
implementing a tax or levy of either type, and we have attempted to
outline 
some key principles that we think need to be recognised and
incorporated in the formulation of a successful long-term strategy.
These include an acknowledgment that 
modern industries have inherited the benefits of our fossil-fuel
heritage (our carbon inheritance), and should also share some of the
cost associated with those past emissions. 
An energy tax or levy offers one simple long-term mechanism 
that is consistent with the principles we have outlined here, and can
be arranged to ensure the most polluting industries pay the most, while 
allowing a smooth transition of costs to low carbon industries as
fossil fuel industries are replaced.  
The insurance industry is uniquely positioned to lead an independent 
or complementary response to climate change, and we encourage
organisations such as The Geneva Association to explore these possibilities.  

\vspace{.5cm}
{\bf Acknowledgments}
\vspace{.5cm}

Richard Clarke's insights about event attribution and insurance
initiated this work, and are explored further in the book ``Predicting
the price of carbon'' \cite{Clarke}.
The notion and importance of ``carbon inheritance'', was identified by   
Anthony Webster, who wrote the paper with comment from Richard
(richard.clarke@icheme.org).
Many thanks to the anonymous reviewers for your time and
valuable suggestions. 

\appendix

\section{}

The tax or insurance levy could be set uniformly for all energy
industries, or determined by an industry's rate of carbon emissions,
with the largest polluters paying the most. 
For example, if $c_i$ is the rate of carbon emissions by an industry
$i$, and $a$ determines the carbon inheritance paid for by all energy
producers, then the fraction of climate change cost allocated to an
industry could be set as $(a+c_i)/\sum_{i}(a+c_{i})$, so that the sum
over all industries is $1$. 
As fossil fuel industries decline, and the amount of carbon they emit
is reduced, then the denominator ($\sum_{i}(a+c_i)$) will become
smaller, and the proportion of climate change cost paid for by (low
carbon) energy producers will smoothly increase, ensuring a steady
long-term revenue to mitigate climate change. 
The cost associated with carbon inheritance can be determined in a
number of ways, such as those discussed in the main text, or through
simple considerations such as the past rate of emissions in
developed countries.   
An option is to increase the proportion paid by fossil
fuel producers and use this extra income to offset the costs of
low-carbon industries, possibly to the extent that they represent a
positive income.  
The intention would be to shift the balance in favour of low-carbon
technologies, so as to accelerate their development. 

\end{document}